\documentclass[preprint2]{aastex}

\usepackage{txfonts}
\usepackage{graphicx}
\usepackage{natbib}
\usepackage{enumerate}

\bibpunct{(}{)}{;}{a}{}{,} % to follow A\&A style

\newcommand{\simgt}{\lower.5ex\hbox{$\;\buildrel>\over\sim\;$}}
\newcommand{\simlt}{\lower.5ex\hbox{$\;\buildrel<\over\sim\;$}}
\newcommand{\hst}{{\sl HST}}
\newcommand{\spit}{{\sl Spitzer}}
                      
\newcommand{\msun}{{$M_\odot$}}                      
\newcommand{\lsun}{{$L_\odot$}}                      
\newcommand{\av}{{$A_V$}}

\newcommand{\zphot}{$z_{\rm phot}$}

\newcommand{\oii}{[O\,{\sc ii}]}
\newcommand{\oiii}{[O\,{\sc iii}]}

\newcommand{\ha}{{H$\alpha$}}                      
 
\newcommand{\lir}{$L_{\rm IR}$}

\newcommand{\grb}{GRB\,080207}

\slugcomment{Final version 20/5/2011}

\shorttitle{The red host of GRB\,080207}
\shortauthors{Hunt et al.}

\begin{document}


\title{The extremely red host galaxy of GRB\,080207}


\author{Leslie Hunt}
\affil{INAF-Osservatorio Astrofisico di Arcetri, Largo E. Fermi 5, I-50125 Firenze, Italy}
\email{hunt@arcetri.astro.it}

\author{Eliana Palazzi}
\affil{INAF-IASF Bologna, Via Gobetti 101, I-40129 Bologna, Italy}

\author{Andrea Rossi}
\affil{Th\"uringer Landessternwarte Tautenburg, Sternwarte 5, D-07778 Tautenburg, Germany}

\author{Sandra Savaglio}
\affil{Max-Planck-Institut f\"ur Extraterrestrische Physik, Giessenbachstra\ss e, 
D-85748 Garching bei M\"unchen, Germany}

\author{Giovanni Cresci}
\affil{INAF-Osservatorio Astrofisico di Arcetri, Largo E. Fermi 5, I-50125 Firenze, Italy}

\author{Sylvio Klose}
\affil{Th\"uringer Landessternwarte Tautenburg, Sternwarte 5, D-07778 Tautenburg, Germany}

%\author{Daniele Malesani}
%\affil{Dark Cosmology Centre, Niels Bohr Institute, University of Copenhagen, Juliane Maries vej 30, 2100 Copenhagen Ø, Denmark} %conflicted since colleague of data proposer

\author{Micha{\l} Micha{\l}owski}
\affil{Scottish Universities Physics Alliance, Institute for Astronomy, University of Edinburgh, Royal Observatory, Edinburgh, EH9 3HJ, UK}

\and

\author{Elena Pian}
\affil{ INAF, Trieste Astronomical Observatory, via G.B. Tiepolo 11, 34143 Trieste, Italy; Scuola Normale Superiore, Piazza dei Cavalieri 7, 56126 Pisa, Italy}



\begin{abstract}
We present optical, near-infrared, and \spit\ IRAC and MIPS observations of the
host galaxy of the dark gamma-ray burst \grb.
The host is faint, with extremely red optical-infrared colors 
($R-K\,=\,6.3$, 24\micron/$R-$band flux $\sim1000$) making it an extremely red
object (ERO) and a dust-obscured galaxy (DOG).
The spectral energy distribution (SED) shows the clear signature of the 1.6\,\micron\
photometric ``bump'', typical of evolved stellar populations. 
We use this bump to establish the photometric redshift \zphot\
as 2.2$^{+0.2}_{-0.3}$, using a vast library of SED templates, including M\,82. 
The star-formation rate (SFR) inferred from the SED fitting
is $\sim$119\msun\,yr$^{-1}$, the stellar mass $3\times10^{11}$\,\msun, and
\av\ extinction from 1-2\,mag.
The ERO and DOG nature of the host galaxy of the dark \grb\ may
be emblematic of a distinct class of dark GRB hosts, with high SFRs, evolved
and metal-rich stellar populations, and significant dust extinction within the host galaxy.
\end{abstract}




\section{Introduction \label{sec:intro}}

Gamma-Ray Bursts (GRBs) without an optical counterpart of their X-ray afterglow are usually called ``dark
bursts''; a more precise definition based on their X-ray to optical spectral energy
distribution (SED) has been proposed
\citep{jakobsson04,rol05,vanderhorst09}. It is clear now that in most
cases extinction by dust in their host galaxies makes these afterglows
optically dim \citep{greiner11}, while cosmological Lyman drop out (high
redshift) or intrinsic faintness are the exception rather than the rule
\citep[][]{cenko09,perley09,rossi11}.

The long-duration \grb\ is one of these truly dark GRBs.
It was detected by {\it Swift}/BAT on 2008~February~7 at 21:30:21 UT.
{\it Swift}/XRT began observing the field 124~s after the BAT trigger and found
a bright X-ray afterglow, with a positional uncertainty radius of
1\farcs4 \citep{racusin08a}.
No optical/near-infrared afterglow was detected for \grb, despite heroic efforts 
by several ground-based observatories. 
%While the abolute deepest limit on the target where reached by GMOS camera
%on the Gemini South telescope ($griz <$ 24.1, 24.5, 24.2, 25.0  at $9.8$ hr, \citealt{Cucchiara08_GCN7276}), 
The deepest 
limit relative to the fading X-ray afterglow was achieved by the Zeiss-600 
telescope at Mt.Terskol observatory which did not detect the afterglow down to $R_{\rm AB}>20.5$ at 1.69 hr
after the burst \citep{andreev08}.
% xxx
% With the X-ray afterglow spectral index $\beta_{\rm X}=$1.50$^{+0.19}_{-0.18}$ and 
%the spectral index of the (prompt) X-ray to optical afterglow emission $\beta_{OX}<0.26$,
%\grb\ is classified as a dark burst according to both the criteria of
\citet{racusin08b,racusin08c} report that the XRT light
curve between 1.30\,hr %4700 
and 4.72\,hr %17000 seconds 
after the GRB start time
declines monotonically with a power-law of index
$\alpha_X=1.85\pm0.10$. Therefore, we have assumed that the optical
light curve in the same time interval is also decreasing
monotonically, so that its behavior is sufficiently regular that the
criterion of \citet{jakobsson04} to define GRB ``darkness'' --
based on the comparison of the X-ray and optical flux levels at 11\,hr --
can be applied to the optical upper limit $R_{\rm AB}>20.5$
measured %at Mt.Terskol 
at 1.69 hr %($=$ 6100 seconds) 
after the trigger. The X-ray flux at 1.69 hr is $\sim$0.35 c/s ($\sim
3.1 \times 10^{-11}$ erg\,s$^{-1}$\,cm$^{2}$), which, assuming the spectral index
fitted by Racusin et al. to the average XRT spectrum in the above
time interval, $\beta_{X}=1.4\pm0.2$, corresponds to a 1\,keV flux 
of 4.2 $\mu$Jy. Together with the simultaneous optical upper limit,
this yields an optical-to-X-ray index of $\beta_{OX} < 0.27$, which
leads to a dark GRB classification, according to
\citet{jakobsson04} and \citet{vanderhorst09}. 

%With no well-defined optical afterglow position, the position of the
%host galaxy of \grb\ was undetermined before the
%release of the {\it Chandra} Source Catalog \citep{evans10}.
% xxx
%Without the sub-arcsecond optical afterglow position,
%Within the XRT error circle, \citet{rossi11} found two very faint host
%candidates, visible in the VIMOS/$R$-band optical image.
% xxx xxx
In the vicinity of the XRT error circle, \citet{rossi11} found two very faint host
candidates, 
one better visible in VIMOS/$R$-band and the second brighter in $K$.
%visible in the VIMOS/$R$-band optical image.
% xxx
However,
the {\it Chandra} satellite observed the field of \grb\ nine days after the burst detection 
by {\it Swift} \citep{racusin08a}, and was able to localize 
the GRB X-ray afterglow with very high accuracy of 0\farcs67 \citep{evans10}:
$\alpha$\,=\,13:50:02.97, $\delta$\,=\,07:30:07.8 (J2000).
%recent work, based on the release of the {\it Chandra}
%source catalogue \citep{evans10}, has allowed the precise (to within
%$\la 1$\arcsec) localization of the afterglow; 
% xxx
% xxx xxx
%\citet{rossi11} found at this position 
This position coincides with one of the candidates found
by \citet{rossi11} (source ``B''): a very faint ($R_{\rm AB}\sim$ 26.5\,mag), 
very red [$R-K\ga$6, $(R-K)_{\rm AB}\sim4.7$] galaxy.
The other candidate is bluer \citep[dubbed ``A'' by ][]{rossi11}, with  $(R-K)_{\rm AB}\sim2.1$, 
and located just outside the 90\% XRT error circle,
$\sim$2\farcs3 northeast of the {\it Chandra} position. 
Hence we associate the first source (B) with \grb.

In this Letter, we present 
proprietary and archival optical and near-infrared (NIR) observations,
combined with archival \spit\ IRAC and MIPS data of the host of \grb.
We estimate the photometric redshift \zphot\ of the host
by fitting the observed SED with a vast library of GRASIL models 
\citep{silva98,iglesias07,michal08,michal10}. 
Despite its optical faintness,
because of the clear photospheric NIR bump at
rest-frame $\lambda$=1.6\,\micron, 
we are able to estimate the redshift of the host
of \grb\ with a precision of $\pm$0.3.
Section \ref{sec:data} describes the data reduction and the methods
used to derive the photometry.
The SED models are presented in $\S$ \ref{sec:sed}, and
we discuss the results and their implications in $\S$ \ref{sec:discussion}.
Throughout the paper, we assume a 
$\Omega_m$\,=\,0.3, $\Omega_\Lambda$\,=\,0.7 cosmology, with 
Hubble constant $H_0$\,=\,70\,km\,s$^{-1}$\,Mpc$^{-1}$.

\section{Data and photometry \label{sec:data}}

In the course of an analysis of GRB host galaxy SEDs (Hunt et al., in preparation),
we culled the \spit\ archive for observations of \grb.
It was observed by A. Levan (PID 50562) more than one year after the burst
in all four IRAC channels \citep{fazioirac} and about five months after the burst
at MIPS 24\,\micron\ \citep{riekemips}.
$R$- and $K$-band data were taken from \citet{rossi11}.
$B$- and $R$-band images were acquired at the Large Binocular Telescope (LBT)
in the course of our approved observing program; 
$i^\prime$ and $z^\prime$ images were taken from the Gemini archive, 
and an $H$-band image from the \hst/NICMOS archive. 
Finally, we retrieved the ESO/SINFONI NIR integral field spectral data 
from the ESO archive, which
provided a broad-band $J$ continuum point. 
Altogether we analyzed 12 photometric data points, 
from 0.4\,\micron\ to 24\,\micron, in order to establish
the photometric redshift and the general properties of the host of \grb.
The magnitudes given below are corrected for foreground Galactic extinction 
assuming $E(B-V)=0.023$ mag \citep{schlegel98}
and a ratio of total-to-selective extinction of $R_V=3.1$.

\subsection{Optical and near-infrared observations \label{sec:obs}}

We observed the field 2 years after the burst with LBT/LBC in the $B$ 
and $R$ bands, and with VLT/VIMOS in $R$ and VLT/ISAAC in $K_s$ bands. 
Optical and NIR data were reduced using standard IRAF 
tasks\footnote{IRAF is distributed by the National Optical Astronomy 
Observatory, which is operated by the Association of Universities for Research in Astronomy 
(AURA) under cooperative agreement with the National Science Foundation.}, 
and analyzed through aperture photometry. %\citep{Tody1993}. 
We used an aperture size twice the Full Width Half Maximum (FWHM) of the stellar PSF.
Optical VLT and LBT images were calibrated using Landolt stars in the field PG1047$+$3.
ISAAC fields were calibrated  using 2MASS field stars. 

Within the {\it Swift}/XRT error circle of 1\farcs4, and coincident with afterglow 
localized by the {\it Chandra} satellite, an extended object (1\farcs6$\times$0\farcs9) 
is visible in our 
LBC, VIMOS $R$-band, and ISAAC $K_s$-band images; it is thus identified as 
the optical counterpart of the GRB host galaxy \citep{rossi11}. 
The galaxy is marginally detected in both LBC and VIMOS 
$R$ band with $R_{\rm AB}\,=\,26.43\pm0.37$, but is clearly visible in our ISAAC images 
(see Fig. \ref{fig:images}: $K_{s, {\rm AB}}\,=\,21.77 \pm 0.14$).
%(Magnitudes are corrected for Galactic extinction; see above.) 
The object is absent in our $B$-band LBT image, which is however compromised by 
worse seeing. 
%resulting in possible contamination by the fuzzy object $\sim2$\arcsec\ to the north.

The host galaxy is also clearly visible in an archival Gemini/GMOS $z^\prime$ image
obtained 6 months after the burst during very good sky conditions, but
not in the $i^\prime$ image taken immediately after the GRB.
%After correcting for Galactic extinction ($E(B-V)$\,=\,0.023, see above) 
We obtain $i^\prime_{\rm AB}>24.6$ and $z^\prime_{\rm AB}=25.02\pm0.02$. 

We reduced archival $J$ (30 min exposure) and $H+K$ 
(45 min) band SINFONI integral field spectra 
around the host galaxy, following all the typical reduction steps 
applied to near-IR spectra. After background subtraction, the data were 
flat-fielded, corrected for dead/hot pixels, wavelength and flux 
calibrated using telluric stars. The resulting datacubes were collapsed 
to obtain broadband continuum images; although the final S/N in the $H+K$ 
band was too low, the derived $J$ band image was analyzed using standard 
aperture photometry to extract the flux.
The spectra themselves were of insufficient sensitivity for 
reliable detection of \oii, \oiii, or \ha\ emission lines (see $\S$\ref{sec:discussion}).

We reduced the \hst/NICMOS F160W (NIC3) image starting with the standard 
flat-fielded calibrated frames after correcting for the pedestal effect. 
%\citep[for a complete description, see][]{hunt04}.
These frames were then ``drizzled'' onto the final mosaic 
%with the IRAF task {\tt drizzle} 
\citep{fruchter02}, using a pixel filling factor of 0.65,
and maintaining the original pixel size of 0\farcs2.
The host galaxy of \grb\ is clearly visible in this image also, and using 
the same methods as described above, we obtain a flux
of 2.9$\pm$0.7\,$\mu$Jy (see Table \ref{tab:fluxes}).

\subsection{\spit\ IRAC and MIPS observations}

For both IRAC and MIPS observations, we 
started the data reduction with the
Basic Calibrated Data (BCD) and the corresponding masks 
(DCE masks), furnished by
the \spit\ Science Center (SSC) pipeline. %(18.7.0 IRAC, 18.13.0 MIPS). 
The image mosaicing and source extraction package \citep[MOPEX,][]{mopex} was
used to co-add BCD frames for each source, and mask out bad pixels. 
%and additional inconsistent pixels were removed by means of the MOPEX outlier 
%rejection algorithms. 
For MIPS-24, we discarded the first two frames of each observation sequence,
as advocated by the MIPS Instrument Handbook. 
Also for MIPS, 
we incorporated the sigma-weighting algorithm since
it gave less noisy MIPS mosaics than without.
The frames were corrected for geometrical distortion and
projected onto a fiducial coordinate system
with pixel sizes of 1\farcs20, roughly equivalent to the original IRAC pixels.
The quality of our IRAC and MIPS maps is substantially better than that provided
by the \spit\ archive.

\begin{figure*}
\vspace{0.2cm}
\centerline{
\includegraphics[angle=0,height=0.3\linewidth]{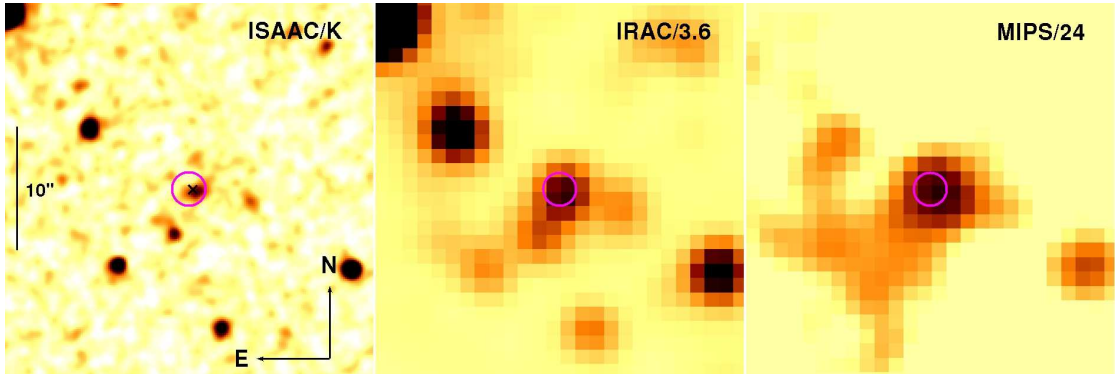}
}
\caption{From left to right: the host galaxy of \grb\ in $K$ band \citep{rossi11},
in IRAC/3.6\,\micron, and MIPS/24\,\micron.
North is up, East is left; the images are 30\,\arcsec\ on a side.
The XRT error circle is shown, and the {\it Chandra} position is marked with $\times$.
\label{fig:images}}
\end{figure*}

Figure \ref{fig:images} shows the ISAAC $K-$band image taken from \citet{rossi11}, 
together with the IRAC 3.6\,\micron\ and MIPS 24\,\micron\ images presented here.
The {\it Chandra} afterglow position is marked with an $\times$.
% xxx xxx
The brightest IRAC pixels correspond to the {\it Chandra} position, and
to source ``B'' found by \citet{rossi11}.

In all IRAC and MIPS \spit\ images we detected only one object within the 
{\it Chandra} error box 
which we identify as the \grb\ host galaxy. The object is not spatially resolved in the 
\spit\ images. 
We estimated the flux density in all filters using a small circular aperture 
with a radius of 2 pixels in IRAC and 3 pixels in MIPS; these measurements were performed 
using the {\tt phot} task in IRAF.
Aperture corrections 
have been applied to account for the extended size of the PSF relative to the small apertures.

Due to the possible contamination from nearby objects (see Fig. \ref{fig:images}), 
we checked the resulting fluxes by 
using PSF fitting with the DAOPHOT software \citep{stetson87}
to perform the source extraction and photometry. 
The results from the two methods are perfectly consistent.
Table \ref{tab:fluxes} reports the fluxes we have derived and use in the
SED fitting described below.

\begin{deluxetable}{ccccc}
\tabletypesize{\scriptsize}
%\rotate
\tablecaption{Photometry of the host galaxy of \grb
\label{tab:fluxes}
}
\tablewidth{0pt}
\tablehead{
 & \multicolumn{1}{c}{($\lambda$)} 
 & \multicolumn{1}{c}{Flux\tablenotemark{a}} 
 & \multicolumn{1}{c}{Uncertainty} 
 & \multicolumn{1}{c}{Telescope/} \\ 
\colhead{Filter} & \colhead{\micron} & \colhead{($\mu$Jy)} & \colhead{$\mu$Jy} & \colhead{Instrument} 
}
\startdata
$B$ & 0.44 &        $<$0.15 \tablenotemark{b} &    & LBT/LBC \\
$R$ & 0.66 &        0.093  &   0.026  & VLT/VIMOS \\
% # 0.66 &        0.054 &   -99
$i^\prime$ & 0.78  &  $<$0.32\tablenotemark{b} &   & Gemini/GMOS \\
$z^\prime$ & 0.925 &  0.35  &    0.06  & Gemini/GMOS \\
$J$ & 1.2   &  1.6   &    0.3    & VLT/SINFONI \\
$H$ & 1.60  &  2.9   &    0.7   & HST/NICMOS \\
$K_s$ & 2.2   &  7.3   &    1.0    & VLT/ISAAC \\
&  3.550 &   14.40   & 0.31  & \spit/IRAC \\
&  4.493 &   15.51   & 0.44  & \spit/IRAC \\
&  5.731 &   18.53   & 1.58  & \spit/IRAC \\
&  7.872 &   12.52   & 1.76  & \spit/IRAC \\
& 23.680 &   92.43   & 6.50  & \spit/MIPS \\
\enddata
%% Text for table notes should follow after the \enddata but before
%% the \end{deluxetable}. Make sure there is at least one \tablenotemark
%% in the table for each \tablenotetext.
\tablenotetext{a}{Blueward of 3.5\,\micron, these values have been corrected 
for Galactic extinction as described in the text.}
\tablenotetext{b}{3$\sigma$ upper limits.}
%with the deluxetable environment}
%\tablenotetext{b}{Another sample footnote for table~\ref{tbl-1}}
\end{deluxetable}

%% If you use the table environment, please indicate horizontal rules using
%% Do not put multiple tabular environments within a single table.
%% The optional \label should appear inside the \caption command.

\section{Modeling the SED \label{sec:sed}}

With $(R-K)_{\rm AB}\sim4.7$ [$(R-K)_{\rm Vega}\sim6.3$], the host galaxy is very red, 
among the reddest host galaxies of a LGRB ever observed.
Such red $R-K$ colors classify the host galaxy of \grb\ as an extremely-red object 
%\citep[ERO: ][]{elston88,wilson04}.
\citep[ERO: ][]{elston88}.
The $R$-24\,\micron\ color is also sufficiently red 
(24\,\micron/$R$ flux ratio $\sim$994) to classify it 
as a dust-obscured galaxy \citep[DOG: 24\,\micron/$R$ flux ratio $\ga$1000,][]{dey08}.

\begin{figure}
\vspace{0.2cm}
\centerline{
%\hbox{\includegraphics[angle=0,bb=40 225 582 640,width=\linewidth]{GRB080207_sedsed_photz.ps}
\hbox{\includegraphics[angle=0,bb=40 225 582 640,width=\linewidth]{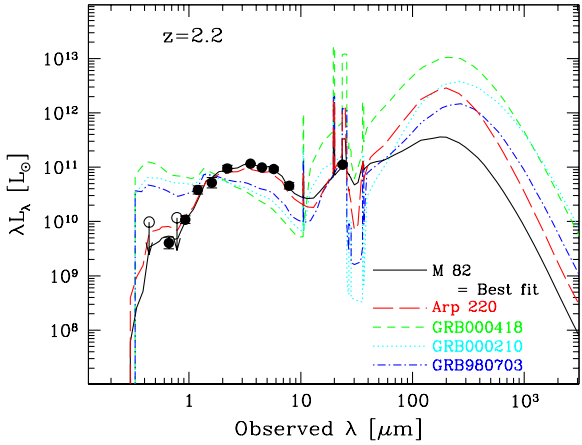}
}
}
\caption{The observed SED (corrected for Galactic extinction as described
in the text) of the host galaxy of \grb\ plotted against observed wavelength.
\label{fig:seds}}
\end{figure}

Given the extremely dim optical fluxes,
the strongest constraint on the SED we present here is the spectral
``bump'' at rest-frame 1.6\,\micron\ (see Fig. \ref{fig:seds}). 
This bump is due to the minimum in the opacity in
the H$^-$ ion found in the photospheres of cool stars, and found
in virtually all stellar populations older than $\sim$100\,Myr; 
it provides a conspicuous
signature with which to constrain photometric redshifts
\citep{simpson99,sawicki02}.
The host of \grb\ is one of the few GRB hosts to show this feature
\citep[another one is ESO\,184$-$G82, the host of GRB\,980425,][]{michal09}. 
%Indeed, the importance of the 1.6\,\micron\ bump for \spit/IRAC colors
%of high-redshift galaxy populations was recognized early on \citep{wright94}; 
%characterizing distant galaxies through
%photometric redshifts with IRAC colors was one of the main design  
%constraints for the instrument.
%
%The presence of this IRAC$+$NIR bump in the SED of the host galaxy of \grb\
%prompted us to consider this constraint for estimating its redshift.
Because of its extremely red colors, 
the host of \grb, is either a
passive early-type galaxy at redshift $1<z<2$,
or a dusty starburst at higher redshift. 
Because LGRBs occur in star-forming galaxies
%\citep[e.g.,][]{lefloch03,fruchter06,savaglio09,castro10},
\citep[e.g.,][]{lefloch03,savaglio09},
we concentrated on the second alternative, and considered %two starburst SEDs
% xxx
a library of $\sim35\,000$ GRASIL star-forming galaxy templates, including M\,82 and Arp\,220
and other GRB host templates \citep{silva98,iglesias07,michal08,michal10}.

\begin{figure}
\vspace{0.2cm}
%\centerline{
%\hbox{
%\includegraphics[angle=0,bb=40 225 582 640,width=\linewidth]{GRB080207_chisq_photz.ps} 
\includegraphics[angle=0,bb=40 225 582 640,width=\linewidth]{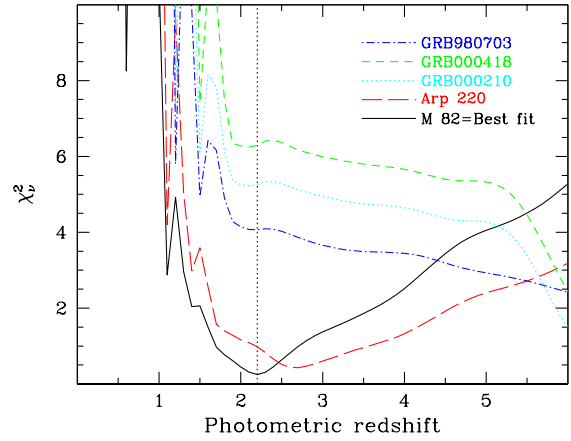} 
%}
%}
\caption{\grb\ host galaxy $\chi^2_\nu$ of the NIR bump plotted against redshift.
The best-fit redshift $z\,=\,2.2$ is indicated by a dotted line.
\label{fig:chisq}}
\end{figure}

SED fitting in $\nu L_\nu$ space was performed by renormalizing the templates,
to minimize $\chi^2$ over the observed SED\footnote{In order to
better weight the NIR bump in the fits, we excluded both
the 8\,\micron\ IRAC point and the 24\,\micron\ MIPS point in the fits;
the fits were more stable with this exclusion.}.
This was done by redshifting the models over a range of redshifts from
0 to 6, in increments of 0.1.
Lyman forest attenuation was treated according to \citet{madau95}.
We used the fitting method described in \citet{iglesias07,michal08,michal10} and one
developed independently by our group; the two approaches give virtually
identical results.
The best fit was obtained for a starburst template, with roughly the same
parameters as M\,82, at a 
redshift $z\,=\,2.2$ (mean residuals 0.1$\,\mu$Jy, reduced $\chi^2_\nu$\,=\,0.3).
% Michal says: I have got z= 2.1 (+0.2 -0.3) perfectly consistent with your results.
The best-fitting SED is shown in Fig. \ref{fig:seds},
and the trends with redshift of $\chi^2_\nu$ for representative
GRASIL templates are shown in Fig. \ref{fig:chisq}.
After a statistical analysis of the $\chi^2$ distributions,
we conclude that the best estimate of the photometric redshift
for the host of \grb\ is $z\,=\,2.2^{+0.2}_{-0.3}$
% xxx
(errors are 1$\sigma$, and were calculated including the choice of template as
a free parameter).

The most salient feature of our SED fit is its ability to well
fit the MIPS point, the $B$ and $i^\prime$ upper limits, 
and the $R$, $z^\prime$, and $J$-band fluxes.
%even though they were not considered in the fitting procedure. 
Even though the 24\,\micron\ data point is not included in our
fitting procedure, it is well predicted by the best-fit SED. %is also consistent with the models,
At $z\,=\,2.2$, MIPS-24
falls roughly on the redshifted polycyclic aromatic hydrocarbon
(PAH) feature, ubiquitous in metal-rich starbursts. %\citep[e.g.,][]{brandl06}.
The faint optical fluxes are clearly subject to
significant extinction; the GRASIL starburst models %for M\,82
predict \av$\sim$1\,mag.
However, this could be a lower limit since
these opacities are integrated over the entire galaxy in the models,
and consider the opacity of the molecular clouds, together with their filling
factor over the galaxy disk.
%Using M\,82 as a local analog,
%empirical measurements of extinction on small spatial scales give much higher values,
%\av$\sim50$\,mag \citep{forster01}.
%The implication is that the host of \grb\ is dusty, 
%both globally and possibly on small local scales, and that the optical
%fluxes are attenuated by dust extinction.
%We explore this point further below.

\section{Discussion and conclusions \label{sec:discussion}}

We can use the GRASIL templates to infer the
total infrared (IR) luminosity (\lir) %(\lir$\sim3.8-4.6\times10^{10}$\,\lsun)
%is $\sim6.5-8$\msun\,yr$^{-1}$ \citep{kennicutt98,bell01,gao04}. 
the star-formation rate (SFR), and the stellar mass ($M_*$). 
From the SED fitting procedure, %the normalization factor implies that
%the host of \grb\ is $\sim$23 times more luminous than M\,82,
%\lir$\sim10^{12}$\,\lsun.
%Assuming that the host is a scaled-up version of M\,82
%would imply a SFR of $\sim$150\msun\,yr$^{-1}$.
we obtain a total \lir$\sim6.8\times10^{11}\pm25$\%\,\lsun,
and SFR $\sim119\pm$25\%\,\msun\,yr$^{-1}$. 
This is roughly an order of magnitude lower than the SFRs implied by the
few detections of GRB hosts in the submillimeter \citep{tanvir04},
but would definitely qualify the host of \grb\ as a dusty luminous
star-forming galaxy.

In the absence of extinction, the inferred SFR is sufficiently high that
both \oii\ and \ha\ would have been detected at $>5\sigma$ with 
the SINFONI spectra. 
The (conservative) 5$\sigma$ upper limits we obtain 
(in $J$ for \oii\ and $K$ for \ha) give a SFR$\sim30-40$\msun\,yr$^{-1}$
\citep[using the SFR calibrations of][]{savaglio09}, 
taking into account the uncertainty in redshift, and the impact of the
sky lines on the noise in the observed spectrum. 
This is a factor of 3 to 4 times lower than the unattenuated SFR inferred 
from \lir\ of the SED analysis; if the latter value is the true one, 
the extinction of the emission lines would be \av$\ga1-2$\,mag,
consistent with the SED fitting. 
There is some indication that the extinction could be even higher around the \grb\ location.
%If no break is assumed in the SED at 1.69 hr after trigger
%(corresponding, for z$\sim$2, to 2000 seconds in rest frame, without
%considering possible relativistic time foreshortening), 
%the extrapolation
%of the X-ray flux to the optical R band 
%(assuming a power-law slope $\beta_X\,=\,1.4$ and a 1\,keV flux
%of $3.11\times10^{-11}$ cts\,s$^{-1}$)
%yields R $\sim$ 12.6, which is 13.9
%magnitudes brighter than the R-band upper limit.  
Assuming there is a spectral break between the optical and X-ray band due to
electron cooling ($\nu_C$), the optical spectral slope soon after the trigger
will be $\beta_{opt} = \beta_X - 0.5$, and the R magnitude is fainter
for higher values of the frequency break.  If 
$\nu_C$ coincides with the lower boundary of the XRT spectral range
(0.3 keV), the extrapolated $R$ band magnitude is  $\sim$15.4, 
5.1 magnitudes brighter than the photometric upper limit ($R_{\rm AB}>20.5$).
The optical flux deficit is
thus significant and, if due to dust extinction, it supports
independently the results we have obtained from the analysis of the galaxy
SED.
% xxx xxx

The implied luminosity and SFR places the %putative 
host galaxy of GRB\,080207 into a subgroup of %(submm-detected) 
GRB host galaxies characterized by
violent star formation and clearly distinct from the majority of the GRB
host galaxy population \citep[c.f.,][]{svensson10}.
A representative
example is the host of GRB\,010222, which is supposed to have $L_{\rm bol}
\approx 4\,\times\, 10^{12} L_\odot$ and a SFR of about 600
\msun\,yr$^{-1}$ \citep{frail02}. Also the submm-detected hosts of GRBs
980329, 980703, 000210 and 000418 belong to this category
\citep{berger03,michal08}, while the host of the first dark
burst, GRB\,970828, only has a comparatively modest SFR of
$24^{+43}_{-14}$\,\msun\,yr$^{-1}$ \citep{lefloch06}.

The stellar mass, $M_*$, of the host of \grb\ inferred from the SED fitting
is $\sim3.2\times10^{11}\pm$25\%\,\msun,
one to two orders of magnitude higher than those
%found for GRB hosts at $z\la$1.5 \citep{savaglio09,castro10}.
found for GRB hosts at $z\la$1.5 \citep[e.g.,][]{savaglio09}.
The SED fit gives an absolute rest-frame $K$ magnitude of the \grb\ host: 
$M_K(\rm AB)\sim-25.2$.
From the correlation between $M_*$ and $M_K$ derived by
\citet{savaglio09}, we infer $M_*\sim4.0\times10^{11}$\,\msun,
consistent within the uncertainties of the value derived from the SED analysis. 

\begin{figure}[h!]
\vspace{0.2cm}
%\centerline{
%\hbox{
%\includegraphics[angle=0,width=\linewidth]{massredshiftERO.eps} 
\includegraphics[angle=0,width=\linewidth]{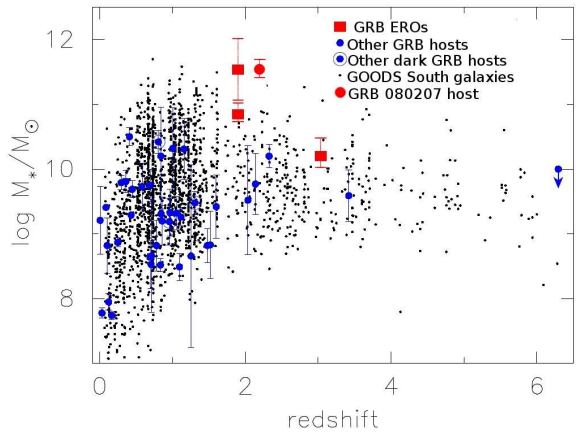} 
%}
%}
\caption{Stellar mass vs. redshift for GRB hosts and other galaxy populations.
Data are taken from \citet{savaglio09}.
\label{fig:masses}}
\end{figure}

Figure \ref{fig:masses} shows stellar masses plotted against redshift for
GRB host galaxies and a set of GOODS galaxies \citep[taken from][]{savaglio09}. 
As can be seen in the figure,
the mass of the host of GRB\,080207 is
comparable to the most massive GRB host galaxies ever measured
\citep[GRBs 020127, 080325:][]{berger07,hashimoto10}.
Besides being quite massive, the hosts of both GRB\,020127 and 
080325 are also EROSs \citep{berger07,hashimoto10},
similar to the host of \grb. 
The SFR and stellar mass of the \grb\ host are 
consistent with the
correlation found for star-forming $BzK$-selected galaxies at $z\,=\,2$
discussed by \citet{daddi07}.
The specific SFR, SSFR, of the \grb\ host is comparable to other galaxy populations at
$z\,=\,2$, with SSFR $\sim 0.4$\,Gyr$^{-1}$.

The metal abundance implied from the mass-metallicity relation
at $z\,=\,2$ \citep{erb06,hayashi09} places the massive host of \grb\ at 
roughly solar metal abundance.
This galaxy is more chemically evolved 
than typical GRB hosts which (at least at lower redshifts) 
tend to be metal poor and much less massive \citep{savaglio09}.
Moreover, the \grb\ galaxy seems to be older than many GRB hosts.
The mass of the stars in the burst is $\la$3\% of the
total stellar mass in the galaxy.
This is consistent with the clear presence of the NIR bump in the SED,
and implies that the mean stellar age is relatively old, $\sim$1-2\,Gyr,
unlike many GRB hosts which tend to be young, $\la$10\,Myr
\citep[e.g.,][]{michal08,levesque10}.

With its extremely red $R-K$ color, $(R-K)_{\rm Vega}\sim6.3$, 
the host of \grb\ is the most extreme ERO 
found to date for a GRB host galaxy\footnote{The host of GRB\,020127 
has $(R-K)_{\rm Vega}\sim6.2$
\citep[at $z\,=\,1.9$:][]{berger07}, and the host
of GRB\,030115 $(R-K)_{\rm Vega}\,=\,6$ \citep[$z\,=\,2.6$:][]{levan06}.}.
%There are also other, less extreme, red GRB host galaxies including 
%GRB\,020819 \citep[$(R-K)_{\rm Vega}\,=\,2.8$, $z\,=\,$0.4:][]{kupcu10},
%GRB\,050122 \citep[$(R-K)_{\rm Vega}\,=\,3.5$, $z\,=\,0.8$:][]{rol07},
%and GRB\,080607 \citep[$(R-K)_{\rm Vega}\,=\,4.0$, $z\,=\,3.0$:][]{chen10}.
Moreover, the unusually high ratio of 24\,\micron-to-$R$-band flux of the \grb\ host
makes it also a dust-obscured galaxy, or DOG \citep{dey08}.
To our knowledge, this is the first GRB host galaxy to be so classified.
Such red host colors at $z\ga$2 stand out from the typical
sub-luminous blue galaxies that tend to dominate lower-$z$ GRB host
populations \citep{savaglio09}. 
In fact the colors and other properties of the host of \grb\ are more typical of
dusty $z=$2 galaxies detected in the IR and submillimeter
\citep{pope08}, star-forming $BzK$-selected galaxies
\citep[e.g.,][]{daddi05,grazian07},
%or Distant Red Galaxies \citep[e.g.,][]{vandokkum04,papovich06}.
or Distant Red Galaxies \citep[e.g.,][]{vandokkum04}.
Indeed, there is more and more evidence 
\citep[e.g., GRBs 020127, 080325, 080607:][]{berger07,hashimoto10,chen10}
that some dark bursts,
those with extremely red optical-IR colors, may
point to a distinct GRB host population: one that harbors more evolved,
more metal enriched stellar populations, 
but with high SFRs and significant dust extinction. 

\acknowledgments

G.C., L.K.H., and E.P. gratefully acknowledge a
financial contribution from the agreement ASI-INAF I/009/10/0.
A.R. and S.K. acknowledge support by DFG
Kl 766/11-3, Kl 766/16-1 and by the German Academic Exchange Service
(Deutscher Akademischer Austausch-Dienst; project D/08/15024). A.R.
acknowledges support from the BLANCEFLOR Boncompagni - Ludovisi, n\'ee Bildt
foundation.


%% in parentheses, after the keyword, but they will not be verified.

{\it Facilities:} \facility{VLT}, \facility{Spitzer}, \facility{LBT}, \facility{HST}, \facility{Gemini}.


\end{document}